\documentclass[twocolumn, twocolappendix]{aastex63}

\newcommand{\y}{\color{red}}
\newcommand{\g}{\color{blue}}

\shorttitle{Day-side Fe\,{\sc i} Emission and Brightness Variation of WASP-33b}
\shortauthors{Herman et al.}

\graphicspath{{./}{}}

\begin{document}

\title{Day-side Fe {\sc i} Emission, Day-Night Brightness Contrast and Phase Offset of the  Exoplanet WASP-33b} 

\correspondingauthor{Miranda K. Herman}
\email{miranda.herman@utoronto.ca}

\author[0000-0003-3636-5450]{Miranda K. Herman}
\affiliation{Astronomy $\&$ Astrophysics, University of Toronto, 50 St. George St., Toronto, ON M5S 3H4, Canada}

\author[0000-0001-6391-9266]{Ernst J. W. de Mooij}
\affiliation{Astrophysics Research Centre, Queen's University Belfast, Belfast BT7 1NN, UK}

\author[0000-0003-4698-6285]{Stevanus K. Nugroho}
\affiliation{Astrophysics Research Centre, Queen's University Belfast, Belfast BT7 1NN, UK}
\affiliation{Astrobiology Center, NINS, 2-21-1 Osawa, Mitaka, Tokyo 181-8588, Japan}
\affiliation{National Astronomical Observatory of Japan, NINS, 2-21-1 Osawa, Mitaka, Tokyo 181-8588, Japan}

\author[0000-0002-9308-2353]{Neale P. Gibson}
\affiliation{School of Physics, Trinity College Dublin, The University of Dublin, Dublin 2, Ireland}

\author[0000-0001-5349-6853]{Ray Jayawardhana}
\affiliation{Department of Astronomy, Cornell University, Ithaca, NY 14853, USA}


\begin{abstract}

We report on Fe {\sc i} in the day-side atmosphere of the ultra-hot Jupiter WASP-33b, providing evidence for a thermal inversion in the presence of an atomic species. We also introduce a new way to constrain the planet's brightness variation throughout its orbit, including its day-night contrast and peak phase offset, using high-resolution Doppler spectroscopy alone. We do so by analyzing high-resolution optical spectra of six arcs of the planet's phase curve, using ESPaDOnS on the Canada-France-Hawaii telescope and HDS on the Subaru telescope. By employing a likelihood mapping technique, we explore the marginalized distributions of parameterized atmospheric models, and detect Fe {\sc i} emission at high significance ($>10.4\sigma$) in our combined data sets, located at $K_{\rm p}=222.1\pm0.4$ km/s and $v_{\rm sys}=-6.5\pm0.3$ km/s. Our values agree with previous reports. By accounting for WASP-33b's brightness variation, we find evidence that its night-side flux is $<10\%$ of the day-side flux and the emission peak is shifted westward of the substellar point, assuming the spectrum is dominated by Fe {\sc i}. Our ESPaDOnS data, which cover phases before and after the secondary eclipse more evenly, weakly constrain the phase offset to $+22\pm12$ degrees. We caution that the derived volume-mixing-ratio depends on our choice of temperature-pressure profile, but note it does not significantly influence our constraints on day-night contrast or phase offset. Finally, we use simulations to illustrate how observations with increased phase coverage and higher signal-to-noise ratios can improve these constraints, showcasing the expanding capabilities of high-resolution Doppler spectroscopy.

\end{abstract}




\section{Introduction}\label{sec:Intro} 

Our ability to probe the atmospheric properties and compositions of exoplanets has progressed rapidly in the last decade, as the relatively young field of exoplanet observation has shifted from an era of discovery to one focusing on characterization. Perhaps the most exciting advance in atmospheric characterization has been the development of high-resolution Doppler spectroscopy \citep[e.g.,][]{Snellen10, Brogi12, Rodler12, Birkby13}. This has significantly improved our ability to detect exoplanet atmospheres using ground-based observations. High-resolution spectrographs allow individual atomic and molecular lines to be resolved, and the large Doppler shift of planetary lines (due to the orbital motion of close-in planets) allows us to disentangle the planetary signal from the relatively stable stellar lines and telluric absorption. By utilizing high-dispersion data and cross-correlation techniques with model templates, the signals from potentially thousands of resolved lines can be combined to boost the planetary signal.

One of the limiting factors in such characterization, however, is the immense brightness contrast between a planet and its host star, which can prevent any atmospheric emission from being detected. Focusing our attention on the hottest and largest planets known, namely hot Jupiters, improves our prospects. 

Hot Jupiters are gas giants that orbit their host stars in a matter of days or even hours, with highly irradiated day-sides as a result of tidal locking. Their extreme environment, paired with a planet's chemical composition, can produce fascinating atmospheric structure, including a stratosphere. Hot Jupiters with effective temperatures $> 1600$ K are expected to harbour such thermal inversions due to the presence of gaseous molecules with strong optical opacities, like TiO and VO \citep{Hubeny03,Burrows07,Fortney08}. These molecules absorb incident stellar radiation, heating the upper atmosphere. 

For hot Jupiters with $T_{\rm eff} > 2500$ K, however, most molecules dissociate and species such as TiO and VO are expected to be less abundant. Instead, other species such as atomic metals, metal hydrides, and H$^-$ opacity may be capable of absorbing enough irradiation to produce a thermal inversion \citep{Lothringer18}. Neutral iron (Fe {\sc i}) is of particular interest as a contributor to thermal inversions in ultra-hot Jupiters. Below 3000 \AA, bound-free transitions absorb high-energy irradiation, while bound-bound transitions absorb significantly above 3000 \AA~ \citep{Sharp07}. As a result, there is an extensive number of Fe {\sc i} emission lines in the optical, and their detection can be used to directly determine whether there is a thermal inversion in the planet's atmosphere. Moreover, given that the abundance of iron is used as a proxy for stellar metallicity, a measurement of atmospheric Fe {\sc i} could potentially lead to a comparison between planetary and stellar metallicities, though meaningful abundance measurements from observed volume mixing ratios are difficult. 

Only recently has Fe {\sc i} been detected in the atmospheres of a handful of hot Jupiters. Using high-resolution spectroscopy, Fe {\sc i} has been found in the transmission spectra of KELT-9b \citep{Hoeijmakers19}, KELT-20b/MASCARA-2b \citep{Nugroho20a, Stangret20, Hoeijmakers20a}, WASP-76b \citep{Ehrenreich20}, and WASP-121b \citep{Gibson20, Cabot20, Hoeijmakers20b}. Additionally, Fe {\sc i} has been detected in emission from the day-side spectra of three ultra-hot Jupiters, KELT-9b \citep{Pino20} WASP-33b \citep{Nugroho20b}, and WASP-189b \citep{Yan20}. These detections not only suggest that atomic species are common in ultra-hot Jupiter atmospheres, but the latter emission detections provide direct evidence of thermal inversions in the absence of TiO or VO. 

It is worth noting that, in some cases, reports of inversion layers have been debated in the literature. The first claim of a thermal inversion \citep[HD 209458b,][]{Knutson08}, for instance, was followed by a handful of discrepant results \citep{Diamond-Lowe14, Zellem14, Schwarz15}. More recently, reports of a TiO-driven thermal inversion in WASP-33b \citep{Haynes15, Nugroho17} have either been questioned \citep{Herman20} or could not be verified convincingly \citep{vonEssen15,Serindag20}. The inconsistencies highlight the necessity of confirming reported detections through independent observation and analysis. We therefore perform an independent search for Fe {\sc i} emission in the day-side of this ultra-hot Jupiter.

\subsection{WASP-33b}
WASP-33b \citep{Collier10} is an ideal target for atmospheric characterization via high-resolution spectroscopy (HRS), given its incredibly high equilibrium temperature \citep[$>3100$ K;][]{Smith11, deMooij13, Haynes15, Zhang18, vonEssen20} and inflated radius \citep[$\sim1.6~R_{\rm J}$;][]{Chakrabarty19}. The planet orbits its host star with a period of $\sim1.22$ days. WASP-33, a fast-rotating $\delta$-Scuti A5 star, is also well-suited for HRS due to its brightness (V = 8.14), though its pulsating nature can introduce complications in the analysis \citep[e.g.,][]{Nugroho20b}.

As mentioned, conflicting results have been found regarding a thermal inversion in WASP-33b when analyzing various data sets in different ways. \citet{Haynes15} first reported evidence of a stratosphere using combined low-resolution transmission data from WFC3/HST and Spitzer, though \citet{vonEssen15} could not rule out an atmospheric model with no inversion based on optical and NIR photometry of the secondary eclipse. \citet{Nugroho17} used HRS to identify TiO in WASP-33b's day-side emission, indicative of a thermal inversion, but this detection was not corroborated by \citet{Herman20}, who used HRS to observe the planet in both transmission and emission. \citet{Serindag20} also reassessed the TiO detection using the original data set of \citet{Nugroho17} and an alternative line list, and found ambiguous results; they report a slightly weaker TiO signal that is unusually offset in both $K_{\rm p}$ and $v_{\rm sys}$, shedding more doubt on the detection. Recently, \citet{Nugroho20b} used their same day-side observations to identify a thermal inversion based on Fe {\sc i} emission, reporting a $6.4\sigma$ signal with $K_{\rm p} = 226$ km/s and $v_{\rm sys} = -3.2$ km/s.

In this work, we present confirmation of Fe {\sc i} emission from multiple day-side observations of WASP-33b using high-resolution Doppler spectroscopy. In Sections \ref{sec:Observation}, \ref{sec:Data}, and \ref{sec:models}, we outline our observations, basic data reductions, and atmospheric emission models, respectively. In Section \ref{sec:Analysis}, we describe our search for Fe {\sc i} using the standard Doppler cross-correlation method, as well as a likelihood-based approach to extract the best-fitting model parameters and their uncertainties. By introducing a brightness variation to this likelihood analysis, we are able to investigate the planet's day-night contrast and peak phase offset with high-resolution spectra for the very first time. 
In Section \ref{sec:Discussion} we present our results, including our measurements of the day-night contrast and phase offset, and discuss their implications, and in Section \ref{sec:sims} we perform the same analysis on simulated spectra to explore the benefits of higher signal-to-noise observations with greater phase coverage. We end with our concluding remarks in Section \ref{sec:conclusion}. The included Appendix provides additional details of our model filtering and plots from our likelihood analyses.

\begin{figure}
	\centering
    \includegraphics[width=0.47\textwidth]{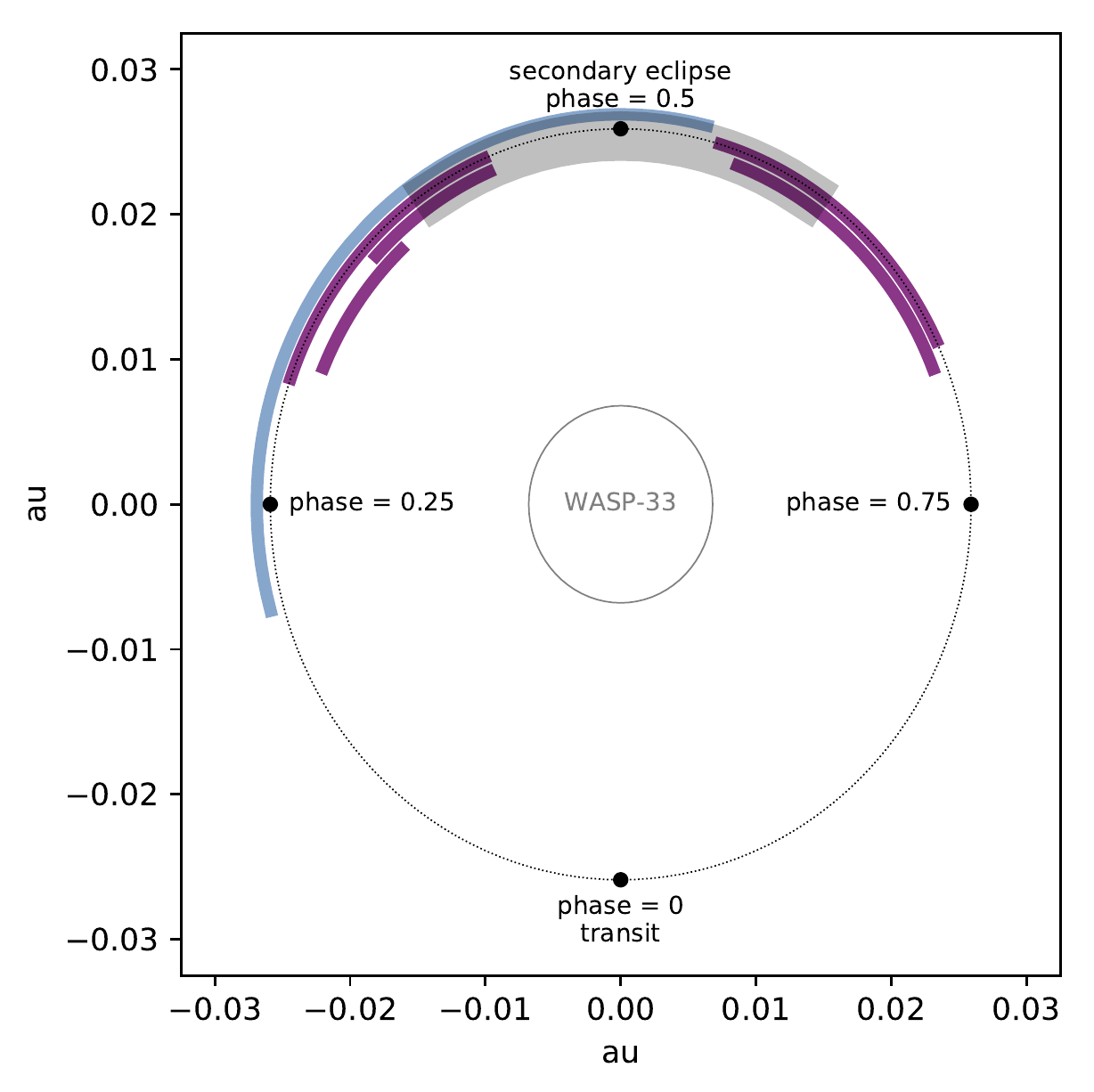}
    \caption{Orbital phases covered during our observations of WASP-33b (Table \ref{tab:obs}). The ESPaDOnS and HDS observations are shown in dark purple and light blue, respectively, and we have shifted the lines radially so that observations with overlapping phases are distinguishable. The grey shaded region indicates the portion of the phase curve we exclude from our analysis due to contamination from stellar pulsations (see Section \ref{sec:croscor}).
    }  
    \label{fig:obs}
\end{figure}

\begin{deluxetable*}{cllccclc}
\tabletypesize{\footnotesize}
\tablecaption{Summary of Observations
\label{tab:obs}}
\tablehead{%
    \colhead{Night} &\colhead{Date} & \colhead{Instrument/Telescope} & \colhead{Duration} & \colhead{Cadence} & \colhead{No. Frames} & \colhead{Orbital Phase} & \colhead{SYSREM Iters}\\
    & \colhead{(UT)} & & \colhead{(hr)} & \colhead{(s)} & & &
    }
\startdata
1 & 2013 Sep 14 & ESPaDOnS/CFHT & 3.9 & 127 & 110 & $0.30 - 0.44$ & 5 \\
2 & 2013 Sep 25 & ESPaDOnS/CFHT & 1.9 & 127 & 55 & $0.37 - 0.44$ & 4 \\
3 & 2014 Sep 3 & ESPaDOnS/CFHT & 3.9 & 128 & 110 & $0.56 - 0.69$ & 4 \\
4 & 2014 Sep 14 & ESPaDOnS/CFHT & 3.9 & 128 & 110 & $0.55 - 0.68$ & 6\\
5 & 2014 Nov 4 & ESPaDOnS/CFHT & 2.0 & 130 & 55 & $0.31 - 0.38$ & 6 \\
6 & 2015 Oct 26 & HDS/Subaru & 8.7 & 600 & 52 & $0.21 - 0.54$ & 6
\enddata
\end{deluxetable*}


\section{Observations} \label{sec:Observation}

We observed WASP-33b on six separate occasions between 2013 and 2015, using two high-resolution spectrographs. In Figure \ref{fig:obs} we show the orbital phases spanned by these observations, and below we summarize the parameters for each instrument. Table \ref{tab:obs} provides additional details for individual nights.

\subsection{ESPaDOnS on CFHT}
Five observations of the day-side of WASP-33b were performed using the Echelle SpectroPolarimetric Device for the Observation of Stars \citep[ESPaDOnS;][]{Donati03} on the Canada-France-Hawaii Telescope (CFHT) between September 2013 and November 2014. These data have been previously presented in \citet{Herman20} and we refer the reader to this work for complete details. To summarize, the data were taken during Queued Service Observing using the `Star$+$Sky' mode, achieving a resolution of $\sim$68,000. We chose an exposure time of 90 s for all nights, with the total number of exposures varying between nights due to weather, observability, and time constraints. Details are given in Table \ref{tab:obs}. The observations cover the $3697 - 10480~\AA$ wavelength range across 40 orders.

\subsection{HDS on Subaru}
A single observation of the day-side of WASP-33b was performed using the High Dispersion Spectrograph \citep[HDS;][]{Noguchi02} on the Subaru telescope on 26 October 2015. These data have previously been presented in \citet{Nugroho17} and \citet{Nugroho20b}, and we refer the reader to these works for complete details. To summarize, the observations used Image Slicer 3 \citep{Tajitsu12} to achieve a spectral resolution of $\sim$165,000, and an exposure time of 600 s with a total of 52 exposures. The observations cover the $6170 - 8817~\AA$ wavelength range across 30 orders, with a gap from $7402 - 7537~\AA$ between the blue and red CCD (containing 18 and 12 orders, respectively).


\section{Data Reduction} \label{sec:Data}

Since our data sets have been previously published, here we provide only an overview of our data reduction procedure. We use the reduced data from HDS \citep[see][]{Nugroho17} and the pipeline-reduced data from ESPaDOnS \citep[see][]{Herman20}. We follow the same procedure for both data sets, as in \citet{Herman20}. For each spectral order, we first apply sigma-clipping to mask outliers three times above the standard deviation of each time series. We then correct for the instrument's blaze response, which is a grating-dependent variation in brightness over time. It also includes other wavelength-dependent effects, such as wavelength-dependent absorption and slit losses. To correct this, we divide each frame by a reference frame (the first frame of the night), bin the frame by 200 pixels, and fit the result with a second-order polynomial, which is then divided out from the full frame. Lastly, we align the extracted spectra to a common wavelength grid in the telluric rest frame by fitting a Gaussian to a spectral line in the O$_2$ $\gamma$-band ($\sim6300~\AA$) for each frame, and shift the spectra to match the wavelength of our reference frame. 

\subsection{Removal of Stellar and Telluric Lines} \label{sec:sysrem}
To remove stellar and telluric features from our spectra, we apply the SYSREM algorithm \citep{Tamuz05}, which has become a standard technique for high-resolution spectroscopy \citep[e.g.,][]{Deibert19,Gibson20,Nugroho20b,Herman20,Turner20}. For low numbers of SYSREM iterations, the planetary signal will not be strongly affected as the planet's radial velocity changes rapidly over the course of the observations, while the stellar and telluric lines remain relatively stable over time.

We use the same SYSREM application as in \citet{Herman20}, performing the iterations on each spectral order separately after converting the flux to magnitudes. For the ESPaDOnS data, we use the pipeline error bars as estimates for the uncertainties, while for the HDS data we use the photon noise. We initially perform 10 iterations on both data sets, and identify the optimal number of iterations for each night by considering the RMS of the data. Generally, after $4-6$ iterations the RMS values in all orders have leveled off and do not decrease noticeably with additional iterations. We list the number of iterations used for each night in Table \ref{tab:obs}. For our cross-correlation analysis, we convert the detrended data back to flux, and subtract a value of one. An example of the full data reduction process applied to a single spectral order is shown in Figure \ref{fig:datareduction_order}. Note that as in \citet{Herman20}, we exclude the first and last three orders of our ESPaDOnS spectra from the following analysis due to a significant reduction in signal-to-noise ratio at these extremes of the wavelength range.


\section{Atmospheric Emission Models} \label{sec:models}

\subsection{Generating Models}
We use the same model emission spectra as \citet{Nugroho20b}, which are available upon request. Here we provide an overview of the models, and direct the reader to \citet{Nugroho20b} for complete details. 

We model the planetary emission spectra assuming a 1D plane-parallel hydrostatic atmosphere, calculated across 70 pressure layers. We adopt the physical planet parameters from \citet{Kovacs13}, where $R_{\rm p}=1.679~R_{\rm J}$ and $M_{\rm p}=3.27~M_{\rm J}$. We then calculate the temperature-pressure (T-P) profile using equation 29 of \citet{Guillot10}, with an intrinsic temperature of 100 K, an equilibrium temperature of 3100 K, a ratio of the mean visible opacity to the mean infrared opacity, $\gamma$, of 2, and a mean infrared opacity of 0.01 cm$^{2}$ g$^{-1}$ (e.g., assuming the infrared regime is dominated by H$^{-}$ opacity). We determine the cross-section of Fe {\sc i} using HELIOS-K \citep{Grimm15} assuming a Voigt profile with natural and thermal broadening, and a line wing cut-off of $10^8$ times the Lorentz line width. We additionally use the line list of \citet{Kurucz18} and the partition function of \citet{Barklem16}.

\begin{figure}
	\centering
    \includegraphics[width=0.47\textwidth]{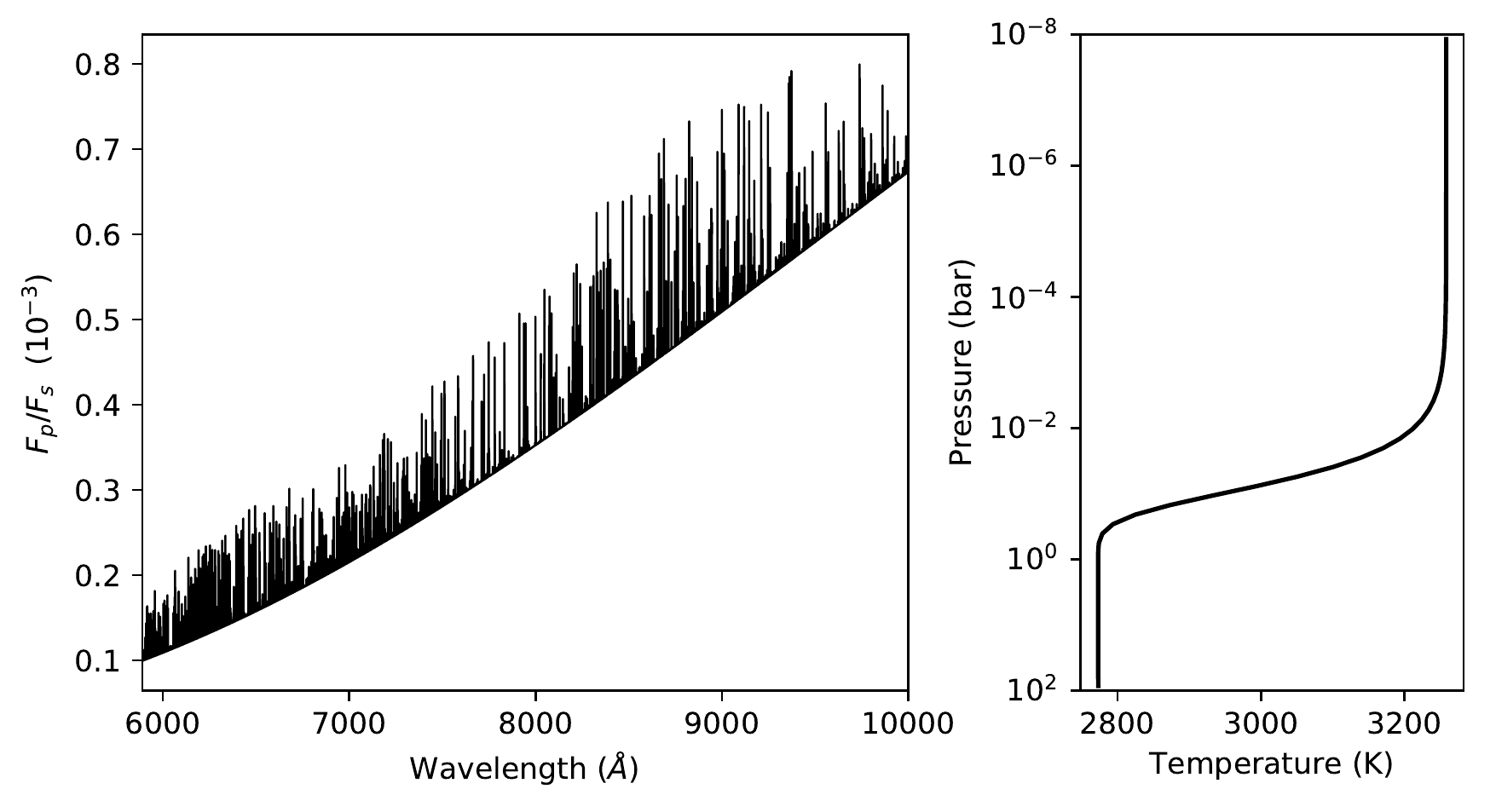}
    \caption{An example of the model Fe {\sc i} emission spectra we use in our analysis with $\log_{10}{\rm VMR}=-4.0$ (left) and our calculated T-P profile (right).}
    \label{fig:model_TP}
\end{figure}

\begin{figure}
	\centering
    \includegraphics[width=0.47\textwidth]{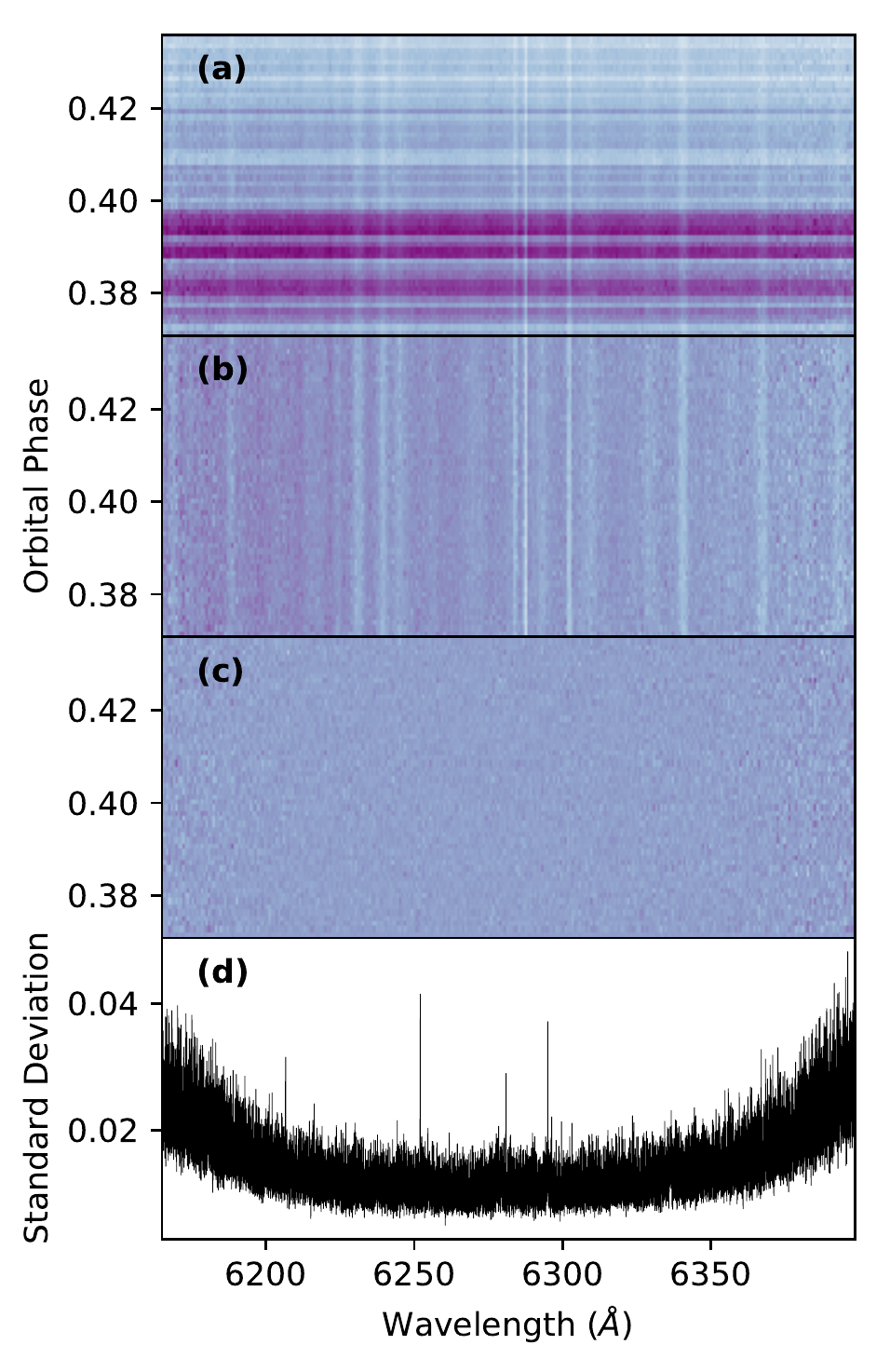}
    \caption{The data reduction process as described in Section \ref{sec:Data}, applied to a single order from our second night of observations with ESPaDOnS.
    \textbf{(a)}: The raw data following the initial reduction pipeline at the telescope.
    \textbf{(b)}: Results of cosmic ray and blaze correction.
    \textbf{(c)}: Resulting spectra after four iterations of SYSREM (Section \ref{sec:sysrem}). 
    \textbf{(d)}: The standard deviation of each data point in the reduced spectra.}
    \label{fig:datareduction_order}
\end{figure}

We include two additional sources of continuum opacity: H- bound-free absorption \citep{John88} and Rayleigh scattering by H$_2$. To estimate the abundances of these species and the mean molecular weight of each atmospheric layer, we use FastChem \citep{Stock18}. We then generate models with a range of Fe {\sc i} volume mixing ratios (VMRs) in log space, from $-5.5$ to $-3.0$ in steps of 0.1 dex. This assumes a constant abundance as a function of altitude. We show a representative example of our model spectra in Figure \ref{fig:model_TP}, alongside the associated T-P profile.

\subsection{Preparing Models for Cross-correlation}\label{sec:modelprep}
To prepare our models for cross-correlation, we convolve each model to the spectral resolution of the instrument using a Gaussian kernel, and calculate the planet-to-star flux ratio by dividing the model by a blackbody spectrum assuming stellar parameters of $R_* = 1.509 ~R_\odot$ and $T_{\rm eff} = 7400$ K. We then apply a high-pass Butterworth filter to the models\footnote{We implement the high-pass Butterworth filter using \texttt{scipy.signal.sosfiltfilt.}}. This is meant to resemble the effect that SYSREM has on the planetary spectral lines within the data, whereby the continuum and some of the line signal is lost in the process of removing stellar and telluric lines (Section \ref{sec:sysrem}). As each night of observations experienced different conditions and therefore requires a different number of SYSREM iterations, the parameters of the Butterworth filter applied to the models necessarily change as well. We discuss how these parameters are determined in Appendix \ref{sec:model_filt}. 

The final planetary spectrum model gives the line contrast relative to the stellar continuum, and this can then be interpolated to the wavelength grid of the data in the following cross-correlation steps.

\section{Analysis} \label{sec:Analysis}

We conduct a search for Fe {\sc i} emission from the day-side of WASP-33b by first pre-processing our data with SYSREM, as discussed in Section \ref{sec:sysrem}, then performing the following Doppler cross-correlation.

\subsection{Doppler Cross-correlation}\label{sec:croscor}
Since our Fe {\sc i} models contain numerous resolved emission lines, cross-correlating our emission spectra with these models allows the signal from many spectral lines to be combined, increasing the significance of the detection. To perform this cross-correlation, each model is Doppler shifted in steps of 1 km/s from $-600$ km/s to $+600$ km/s, interpolated onto the same wavelength grid as the data, and then multiplied by our residual spectra, where each data point is weighted by its variance. We then sum over all wavelengths to produce a cross-correlation function (CCF) dependent on both time and velocity. This process is repeated for each spectral order separately, and all orders are then summed together.

We then shift this CCF to the rest frame of the planet, where the planet's radial velocity at a given orbital phase $\phi$ (assuming a circular orbit) can be expressed as:

\begin{equation}
    RV = K_{\rm p} \sin(2\pi \phi) + v_{\rm sys} + v_{\rm bary}.
\end{equation}
$K_{\rm p}$ is the orbital velocity of the planet, $v_{\rm sys}$ is the systemic velocity of the star-planet system, and $v_{\rm bary}$ is the barycentric correction. We assume a range of orbital and systemic velocities, for a $K_{\rm p}$ of 175 to 325 km/s and a $v_{\rm sys}$ of $-150$ to $+350$ km/s, both in steps of 0.25 km/s. After each frame of the CCF has been shifted to the planet's rest frame via linear interpolation, the CCF is summed over time. This produces a $K_{\rm p} - v_{\rm sys}$ map for each night of observation, which can then be summed together.

However, our CCFs are noticeably impacted by the pulsating nature of the host star WASP-33, which is a $\delta$-Scuti variable. These pulsations cause the stellar line profiles to become distorted as a function of time, and since Fe {\sc i} is also present in the stellar spectrum, cross-correlating with an Fe {\sc i} model produces an unwanted stellar signal in addition to any planetary signal. If this pulsation signal is included in the summation of the CCF over time, it can produce a signal that could be mistaken for a planetary detection. Luckily, the pulsations are limited to a smaller velocity range in our CCFs than the expected planetary signal at most orbital phases. The stellar and planetary signals in the CCF only begin to overlap near secondary eclipse (and transit), meaning the pulsations can be avoided by limiting the phase range included in our summation over time. We therefore follow the example of \citet{Nugroho20b}, only summing up the CCFs outside of the phase range 0.41 to 0.59. 

After this procedure, a clear peak at the expected orbital and systemic velocities can be seen in the resulting $K_{\rm p} - v_{\rm sys}$ map. We may estimate the signal-to-noise (S/N) by dividing this map by its standard deviation, which is calculated by excluding the region $K_{\rm p} < 275$ km/s and $v_{\rm sys} < +50$ km/s so as to avoid the planetary signal. As noted by \citet{Gibson20}, this S/N can provide only a rough estimate of the detection significance, as the exact significance will vary slightly, depending on the chosen noise region.

We therefore provide a second estimate of the S/N following the method used in \citet{Herman20}. For each night, we randomly assign a frame to each phase value, then perform the cross-correlation and summation over time. We repeat this procedure 10,000 times, and use the 15.9 and 84.1 percentiles of the resulting distribution to estimate the noise. The strength of the detected signal can then be compared to this to estimate the S/N.

\subsection{Likelihood Mapping and Brightness Variation}\label{sec:likelihood}
In addition to cross-correlating our spectra with atmospheric models using the standard approach introduced by \citet{Snellen10}, we also compute the likelihood map, following the approach of \citet{Brogi19} and \citet{Gibson20}. The latter allows different models to be explored and compared using principled statistical techniques, and can be calculated directly from the cross-correlation map without repeatedly performing the cross-correlation. We employ the generalized form of the likelihood map from \citet{Gibson20}, which is derived using a similar approach to \citet{Brogi19}, but accounting for both time- and wavelength-dependent uncertainties:

\begin{equation}
    \ln \mathcal{L} = -\frac{N}{2} \ln \frac{\chi^2}{N} 
    \label{eq:logL}
\end{equation}
where $\mathcal{L}$ is the likelihood, $N$ is the total number of data points, and  $\chi^2$ is given by

\begin{equation}
    \chi^2 = \sum \frac{f_i^2}{\sigma_i^2} + A_{\rm p}(\phi)^2 \sum \frac{m_i^2}{\sigma_i^2} - 2 A_{\rm p}(\phi)~ {\rm CCF}.
    \label{eq:chi2}
\end{equation}
As in the CCF, each sum is performed over wavelength. Here $f_i$ is the mean-subtracted spectrum; $\sigma_i$ is the outer product of the standard deviation of each wavelength and exposure bin, normalized by the standard deviation of the spectra in each order; $m_i$ is the mean-subtracted, Doppler-shifted model; and $A_{\rm p}(\phi)$ is a phase-dependent term we introduce to account for the planet's brightness variation throughout the observations, assuming the planetary emission spectrum is dominated by Fe {\sc i} lines. This takes the form:

\begin{equation}
    A_{\rm p}(\phi) = \alpha \big( 1 - C \cos^2(\pi(\phi - \theta)) \big),
    \label{eq:Ap}
\end{equation}
where $\alpha$ is an overall scale factor to determine the strength of the planetary signal relative to the model (i.e., to account for any uncertainty in the scale of the model); $C$ is the day-night contrast in the form $1 - (F_n / F_d)$, where $F_i$ is the observed planetary Fe {\sc i} line emission from the day/night-side measured over the wavelength range of our observations; and $\theta$ is the phase offset of the peak brightness amplitude. Note that the day-night contrast is a bounded parameter limited to values between zero and one, and the phase offset is defined such that a positive offset corresponds to a peak occurring after secondary eclipse. 

We recognize that equation \ref{eq:Ap} may not perfectly match the planet's true brightness variation, and note that more complex models could be explored in the future \citep[e.g.,][]{Cowan11}. We chose this simpler representation largely as a proof of concept: By introducing this variable brightness term, we may shed light on energy re-circulation within the planet's atmosphere. These sort of parameters have typically only been accessible through photometric phase curve measurements \citep[e.g.,][]{Esteves15,Zhang18,vonEssen20}, but we show that they can be constrained using high-resolution Doppler spectroscopy as well. 

We note, however, that we are specifically constraining the phase offset and day-night contrast of the Fe {\sc i} signal, which we assume is the dominant emission source over our wavelength range. While this dependence on Fe {\sc i} may seem like a limitation, it actually demonstrates the power of this technique: By targeting chemical species with different opacities, we may reveal variations in the phase offset and day-night contrast for different molecules/atoms, exploring the atmospheric dynamics at play at different pressure levels. This has a significant advantage over low-resolution spectroscopy, since we will be far less affected by overlapping spectral features, and in the case of WASP-33b, we can minimize the impact of stellar pulsations. We also note that the day-night contrast will predominantly be driven by the change in temperature  in the region probed by the Fe {\sc i} emission lines - since we are operating in the Wien limit, a small change in temperature can significantly impact the flux. The contrast could also be affected by high-temperature cloud coverage, if any, and the condensation of Fe. Of course, a detailed interpretation of $C$ would need to take into account a variable lapse rate as a function of position on the planet. The model we are using is equivalent to many photometric phase-curves, where mostly uniform day- and night-side brightness are assumed, and the variation is caused by the amount of day- and night-side visible.

Returning to the likelihood mapping, we note that unlike the standard definition of the CCF, the CCF in equation \ref{eq:logL} is not weighted/normalized. As \citet{Brogi19} point out, weighting the CCF is no longer required here, as the likelihood contains the data and model variances and thus intrinsically takes the variable S/N of the observations into account.

We compute the log likelihood for $\alpha$ values of 0.5 to 5 in steps of 0.05, day-night contrasts of 0 to 1 in steps of 0.05, and phase offsets of $-30^o$ to $+60^o$ in steps of $1^o$. To do so, the second and third terms in equation \ref{eq:chi2} are shifted to the rest frame of the planet (assuming a range of velocities as is done for the CCF), then all terms in equation \ref{eq:chi2} are added before summing over time and inserting the result into equation \ref{eq:logL}. We note that when calculating the log likelihood for each order, $N$ should be equal to the number of pixels multiplied by the number of exposures. The value of $N$ can differ between orders and nights, and if it is not properly computed, the reduced $\chi^2$ will not be correct, which can impact the resulting best-fit values.

After computing the log likelihood for each spectral order, we take the sum over all orders and all nights. The result is a six-dimensional log likelihood map, dependent on $\log_{10}$VMR, $K_{\rm p}$, $v_{\rm sys}$, $\alpha$, $C$, and $\theta$. To produce the likelihood map from this log likelihood, we first subtract the global maximum, which occurs at the expected $K_{\rm p}$, and $v_{\rm sys}$; this is equivalent to normalizing the maximum likelihood to one. We then compute the exponential. 

This likelihood mapping technique has the advantage over standard CCF maps in that we are now able to directly compare a range of models, and extract the best-fit parameters and their uncertainties. This can be done by exploring slices of the likelihood map (2D conditional distributions) or sums over various parameters (1D marginalized distributions). To estimate the significance of a detection using this method, we determine the median value of the marginalized likelihood of $\alpha$ and divide by its standard deviation.


\section{Results and Discussion} \label{sec:Discussion}
We perform the cross-correlation and likelihood mapping described above for our two data sets (ESPaDOnS and HDS), and consider both the separate and combined results. In all cases, we detect day-side Fe {\sc i} emission at a high significance, confirming the signal reported by \citet{Nugroho20b}. We discuss the details of our constraints in Section \ref{sec:results_data}, but first describe the limitations of our analysis below.

\subsection{Dependence of Fe {\sc i} Emission on Wavelength} \label{sec:bluewaves}
Interestingly, we find that if spectral orders blueward of $\sim 6000~\AA$ are included in our analysis of the ESPaDOnS data, the Fe {\sc i} signal is not detectable at a significant level. This has a relatively straightforward explanation: Below this wavelength, the line contrast is simply too low to be detected, even with multiple observations combined. 

In addition, at the bluest wavelengths our observations are further from the peak of the planet's emission spectrum, meaning even a small change in the temperature of our model could have quite a significant impact on the detected signal. Any underestimation of scattering processes in the atmosphere would also preferentially affect bluer wavelengths.

We verified this low line contrast explanation with simulated data, in which we injected a model with $\log_{10}{\rm VMR} = -3$ and no brightness variation into pure white noise, with the same standard deviation, wavelength range, and phase range as that of each order in the reduced spectra for our first night of ESPaDOnS observations. We Doppler shift the model to the planet's velocity in each frame, interpolate to the wavelength grid of ESPaDOnS, and multiply the model in via: white noise $\times~ (1 + F_p/F_s)$. We then cross-correlated this simulated data with the same model, following the procedure of Section \ref{sec:croscor} for each order. Finally, we compared the S/N of all blue orders combined ($\lesssim6000 \AA$) and all red orders combined ($\gtrsim6000 \AA$). 

We found that the S/N of the injected signal in the blue orders was nearly an order of magnitude lower than that in the red orders. This result confirms that the line contrast at bluer wavelengths is simply too low relative to the noise in the ESPaDOnS spectra to produce any significant detection. By including these orders where the line-contrast is below the detection limit, we are adding noise to the overall signal when all orders are combined. We therefore exclude the orders below $\sim 6000~\AA$ (orders $1 - 24$) in the cross-correlation and likelihood mapping results presented below. The HDS observations only span wavelengths of $6170 - 8817~\AA$, meaning these data are less affected by the decreased line contrast at bluer wavelengths. 

This cutoff choice of $6000~\AA$ is largely a balance between the number of Fe {\sc i} lines available and our wavelength coverage, since eventually the loss of additional lines will instead reduce the significance of our detection. As well, this limit allows us to match the blue ends between HDS and ESPaDOnS. In Section \ref{sec:alt_analysis} we return to this discussion of loss of signal at blue wavelengths, exploring the impact of an alternative temperature structure for our atmospheric models.

\begin{deluxetable}{llcc}
\tabletypesize{\footnotesize}
\tablecaption{Results of the Likelihood Analysis
\label{tab:results_data}}
\tablehead{%
    \colhead{Data Set} &\colhead{Parameter} & \colhead{T-P Profile 1} & \colhead{T-P Profile 2}
    }
\startdata
ESPaDOnS & $K_{\rm p}$ (km/s) & $224.1_{-0.5}^{+0.6}$ & $224.0\pm0.5$ \\
         & $v_{\rm sys}$ (km/s) & $-5.0\pm0.4$ & $-4.6\pm0.4$ \\
         & $C$ & $> 0.90$ & $> 0.90$ \\
         & $\theta ~(^o)$ & $+22\pm12$ & $+18\pm10$ \\
         & $\log_{10}$VMR & $-4.1\pm 0.2$ & $-5.3\pm0.2$ \\
         & $\alpha$ & $3.1_{-0.4}^{+0.3}$ & $1.0\pm0.1$ \\
         & Significance ($\sigma$) & $>9.1$ & $>9.8$ \\
         & & & \\
HDS & $K_{\rm p}$ (km/s) & $223.8\pm3.4$ & $224.6_{-2.2}^{+2.1}$ \\
    & $v_{\rm sys}$ (km/s) & $-4.1\pm2.4$ & $-5.7_{-1.6}^{+1.7}$ \\
    & $C$ & $> 0.97$ & $>0.97$ \\
    & $\theta ~(^o)$\tablenotemark{$*$} & $+44\pm6$ & $+40\pm6$ \\
    & $\log_{10}$VMR & $-4.2\pm0.2$ & $-5.2\pm0.2$ \\
    & $\alpha$ & $3.6_{-0.4}^{+0.3}$ & $1.1\pm0.2$ \\
    & Significance ($\sigma$) & $>8.9$ & $>7.6$ \\
    & & & \\
Combined & $K_{\rm p}$ (km/s) & $222.1\pm0.4$ & $222.1_{-0.5}^{+0.4}$ \\
         & $v_{\rm sys}$ (km/s) & $-6.5\pm 0.3$ & $-5.9\pm0.3$ \\
         & $C$ & $> 0.97$ & $>0.97$ \\
         & $\theta ~(^o)$\tablenotemark{$*$} & $+42\pm6$ & $+38\pm6$ \\
         & $\log_{10}$VMR & $-4.1\pm 0.2$ & $-5.4\pm0.1$ \\
         & $\alpha$ & $3.1\pm 0.3$ & $0.9\pm0.1$ \\
         & Significance ($\sigma$) & $>10.4$ & $>10.2$
\enddata
\tablecomments{We report the median value $\pm1\sigma$ for the marginalized distributions, with the exception of $C$. For this parameter we report the lower limit, determined from the 84th percentile of the distribution.}
\tablenotetext{$*$}{Due to the limited phase range of the HDS data, this value is likely biased and the uncertainties underestimated (see discussion in Section \ref{sec:sims_results}).}

\end{deluxetable}

\subsection{Detection of Day-side Fe {\sc i} Emission and Measurements of Brightness Variation} \label{sec:results_data}
\begin{figure*}
	\centering
    \includegraphics[width=0.66\textwidth]{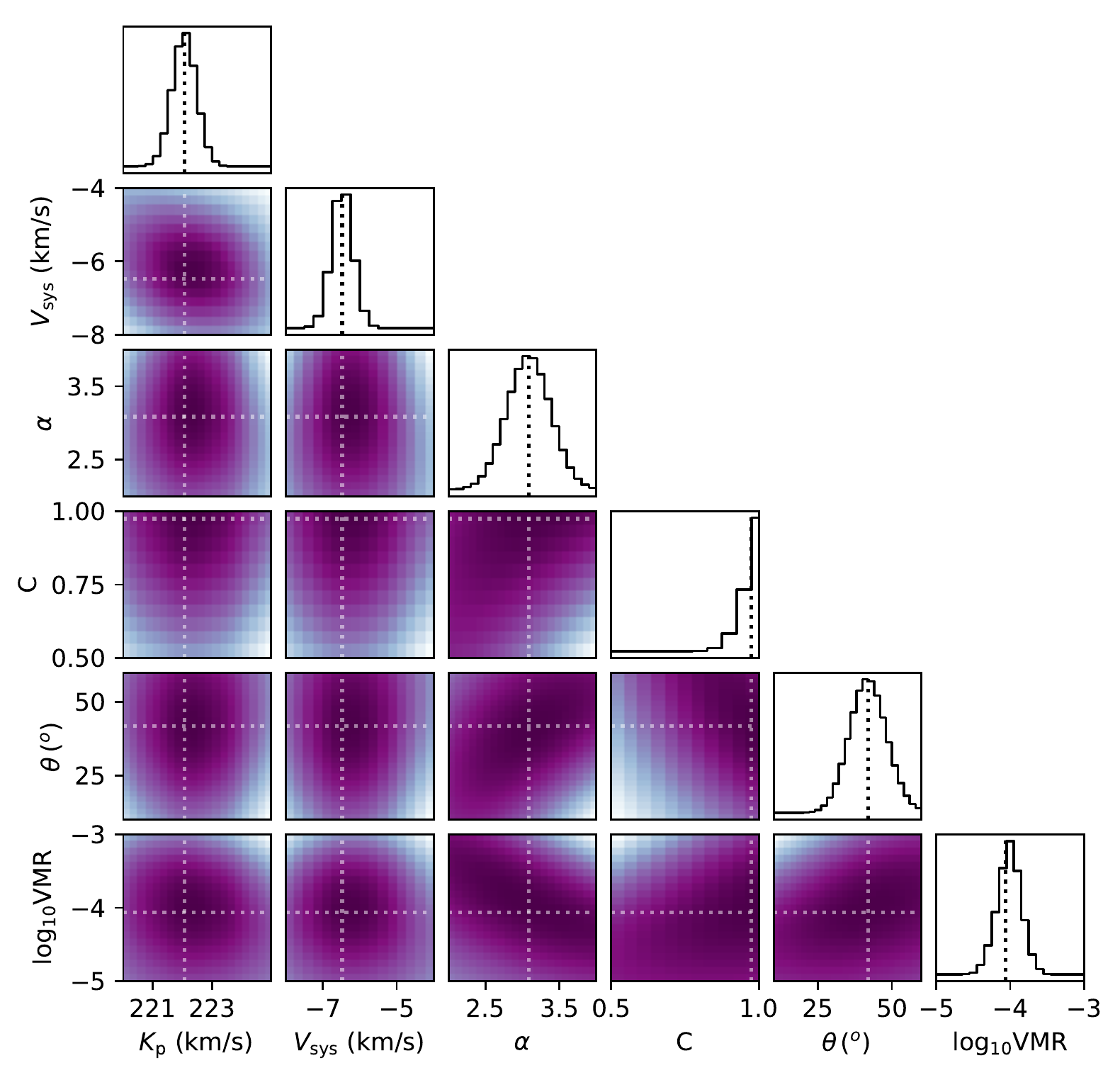}
    \caption{The conditional (2D) and marginalized (1D) likelihood distributions for our combined ESPaDOnS and HDS results, using models with the T-P profile from Figure \ref{fig:model_TP}. The dotted lines shown the median values for each distribution, with the exception of $C$, which shows the lower limit estimated from the 84th percentile of the distribution.
    }
    \label{fig:cornerplot_data}
\end{figure*}

The results of our likelihood mapping are summarized in Table \ref{tab:results_data}, where we present the marginalized parameters for our ESPaDOnS and HDS data sets both separately and combined. For each parameter we report the median value $\pm1\sigma$, with the exception of the day-night contrast $C$. Because this distribution is bounded, we report only the lower limit, determined from the 84th percentile of the distribution. In the following sub-sections we discuss the results from the column labeled `T-P Profile 1', which uses the atmospheric models described in Section \ref{sec:models}. In Section \ref{sec:alt_analysis} we describe the results from the column labeled `T-P Profile 2', where we use models with an alternative temperature structure, for comparison. In Figure \ref{fig:cornerplot_data} we show both the marginalized and conditional distributions for our combined data with T-P Profile 1, and the same figures for the individual data sets can be found in Appendix \ref{sec:extra}. 

In Figure \ref{fig:fmap_data} we show the $K_{\rm p} - v_{\rm sys}$ map for the same combined data. It displays the S/N calculated from the standard deviation of the map, which is not the same as the significance calculated from the marginalized distribution of the $\alpha$ parameter. The former can only provide a rough estimate of the detection significance, as it will depend on which part of the map is used to calculate the standard deviation. We find a S/N of 9.0 based on our $K_{\rm p} - v_{\rm sys}$ map, a S/N of 10.1 based on the phase scrambling method described in Section \ref{sec:croscor}, and a significance of $10.4\sigma$ based on the marginalized distribution of $\alpha$. Thus we have made a strong detection of day-side Fe {\sc i} emission using both standard cross-correlation and likelihood mapping with a brightness variation term.

\begin{figure*}
	\centering
    \includegraphics[width=0.98\textwidth]{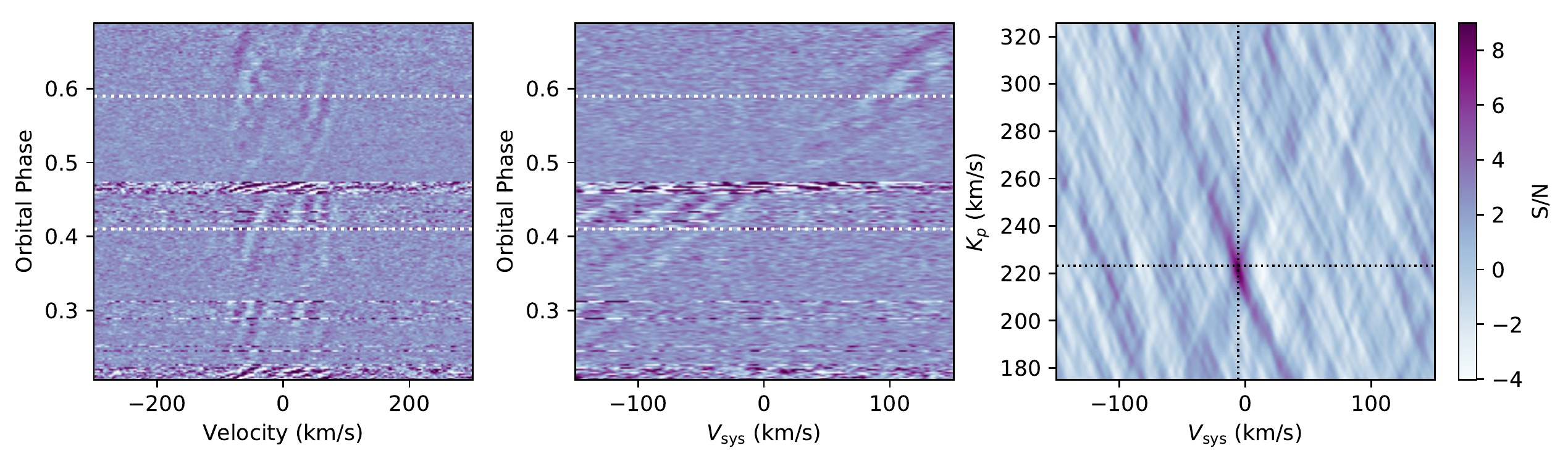}
    \caption{The results of our standard Doppler cross-correlation analysis for our combined data sets, using models with the T-P profile from Figure \ref{fig:model_TP}. The left panel shows the cross-correlation map, where the planet's radial velocity trail is visible as a curved diagonal line starting from the upper left to the lower right. The stellar pulsations are also visible. The region between the white horizontal lines is not included in our later analysis, so as to avoid regions where the pulsations and planetary signal begin to overlap. The center panel shows the same cross-correlation map shifted to the planet's velocity, where the planetary signal now forms a straight vertical line. The right panel shows the $K_{\rm p}-v_{\rm sys}$ map, where we have avoided summing over the orbital phases contaminated by stellar pulsations. The black lines intersect at the map's peak ($K_{\rm p}= 222.1$ km/s and $v_{\rm sys} = -6.5$ km/s), which has a S/N of 9.0. 
    }
    \label{fig:fmap_data}
\end{figure*}

\subsubsection{Velocity Constraints}
In terms of the RV semi-amplitude $K_{\rm p}$ and systemic velocity $v_{\rm sys}$, the values we report in Table \ref{tab:results_data} are in good agreement between the two data sets, and with previously reported values. We note that the HDS results have larger uncertainties than our ESPaDOnS observations because the latter cover both sides of the planet's secondary eclipse, allowing for tighter constraints on velocities.

\subsubsection{Day-Night Contrast}
The lower limits we report for the day-night contrast are also sensible. The most recent phase curve measurements of WASP-33b from the Transiting Exoplanet Survey Satellite (TESS) suggest a day-side temperature of 3014 K and night-side temperature of 1605 K \citep{vonEssen20}, while measurements with $3.6\mu$m Spitzer data find an average day-side temperature of 3114 K and night-side temperature of 1757 K \citep{Zhang18}. 
To provide a simple estimate of the day-night contrast from these measurements based on our description below equation \ref{eq:Ap}, we integrate the Planck function in the Wien's limit, evaluated over the wavelength range of our spectra\footnote{This is a significant simplification, as the spectral response of the instrument is also wavelength dependent. However, integrating over the efficiency of the instrument is complicated, given that there are numerous overlapping orders in both ESPaDOnS and HDS.}. In both cases we arrive at a day-night contrast of $C \simeq 0.99$. Based on the marginalized distribution of $C$ shown in Figure \ref{fig:cornerplot_data}, our data match this result quite well. We emphasize that constraints on bounded distributions are more difficult, however. Since $C$ is hard-bounded at 1, this will will bias recovered values away from 1. This is why we opt to report only the estimated lower limit of the day-night contrast. Even our more conservative lower limit from ESPaDOnS indicates that the night-side flux is at least 10 times less than that of the day-side flux in the wavelength range we consider, assuming Fe {\sc i} is the dominant emission source. In terms of temperature, this contrast of 0.9 indicates that the night-side of WASP-33b is $\lesssim2200$ K, given that our atmospheric models assume a day-side temperature of 3100 K.

\subsubsection{Phase Offset}
Previous constraints of WASP-33b's phase offset, on the other hand, have disagreed with each other significantly. \citet{vonEssen20} find a phase offset of $+28.7^o$ westward of the substellar point, which is at odds with the eastward offset of $-12.8^o$ reported by \citet{Zhang18}. \citet{vonEssen20} suggest that this discrepancy could be due to their analysis of systematics in the data, or the effect of the host star's variability, which plagues the entirety of photometric observations. However, our likelihood analysis also indicates a slight preference for a westward phase offset. And since we exclude phases where the radial velocity of the planet is comparable to that of its host star, our analyses are not impacted by stellar pulsations. 

Notably, our constraints on the phase offset differ considerably between the data sets we present. This can easily be explained by the limited phase coverage of our HDS observations. If the peak brightness occurs after the secondary eclipse, it is difficult to constrain the phase at which this peak occurs when observations are only made before the secondary eclipse. We therefore expect the ESPaDOnS observations, which cover both sides of the day-side phase curve, to more accurately constrain the phase offset. Indeed, our ESPaDOnS constraint of $\theta = +22\pm12$ degrees is comfortably within the uncertainty of the $+28.7\pm7.1$ degrees reported by \citet{vonEssen20}, even though we do not directly observe the phase at which the peak brightness occurs\footnote{A phase range of $0.41-0.59$ would contain a peak with a phase-offset between $-32.4^o$ and $+32.4^o$.}. However, our uncertainty is large enough that our constraint is also within $2\sigma$ of zero offset, and is within $2.9\sigma$ of the eastward offset reported by \citet{Zhang18}. As mentioned previously, our constraint also assumes that Fe {\sc i} emission dominates the planetary flux in our considered wavelength range. We caution directly interpreting this offset as an offset only in the planet's hot-spot location, as our observations are mainly sensitive to the contrast between the lines and the continuum, and so variations in abundance and T-P profile could move the location of the peak from the Fe emission away from the hottest point on the planet.

The constraint from our combined data sets is much closer to our constraint from HDS, because our HDS observations have a higher signal-to-noise ratio (SNR, not to be confused with the detection S/N described in Section \ref{sec:croscor}) than all five of our ESPaDOnS visits combined. We further explore the effect of phase coverage and SNR on our ability to constrain the phase offset and day-night contrast in Section \ref{sec:sims}.

\subsubsection{Fe {\sc i} Abundance}
Regarding the Fe {\sc i} abundance in $\log_{10}$VMR, our constraints are in good agreement with \citet{Nugroho20b}, as expected based on our shared models. These findings suggest that the Fe {\sc i} signal is in emission, as the planet's atmosphere must experience a thermal inversion at the associated pressures \citep[see Figure 4 of][]{Nugroho20b}. As previously mentioned, this thermal inversion has been attributed to both Fe {\sc i} \citep{Nugroho20b} and TiO \citep{Haynes15, Nugroho17}, but given that TiO could not be detected by \citet{Herman20} and its existence was debated by \citet{Serindag20}, we question whether both species could be responsible for the thermal inversion simultaneously. Moreover, the day-side atmosphere of WASP-33b may exceed the temperature at which TiO is thought to dissociate ($T_{\rm eff} > 2500$ K), or at the very least be less abundant, based on the findings of \citet{Lothringer18}. We conclude that the high-significance detections of neutral atomic iron in all data sets presented here provide ample evidence for a thermally-inverted atmosphere in this ultra-hot Jupiter, though we cannot definitively say that Fe {\sc i} is the cause of the thermal inversion without more detailed modeling.

We also emphasize that there is a strong degeneracy between chemical abundance and lapse rate \citep[$dT /d \log P$,][]{Madhusudhan09}. An increase in lapse rate can be compensated for by a decrease in abundance, and vice versa, and so we caution against claims of absolute abundance. The abundance constraints we present here are only valid for the specific T-P profile we consider in our atmospheric models.

\subsubsection{Line Contrast}
The recovered $\alpha$ value in all cases is higher than the constraint from \citet{Nugroho20b}, despite the fact that we use the same planetary emission models. This is because the previous work does not take into account the planet's expected brightness variation as a function of phase, which decreases the line contrast at phases further from the secondary eclipse. Our constraints may also be higher due to the different filtering method we apply to our models (see Appendix \ref{sec:model_filt}). Additionally, our constraints suggest a line contrast that is slightly higher than that expected based on the measured secondary eclipse depth from \citet{vonEssen20}. However, the line contrast is wavelength-dependent (as mentioned in Section \ref{sec:bluewaves}), meaning our scaling term may also have some wavelength dependence that we do not account for. Investigating the issue further is beyond the scope of the current paper, but it would be informative to explore a wavelength- or order-dependent $\alpha$ parameter in future studies.

In any case, our constraints on $\alpha$ still imply that our models have underestimated the strength of the line contrast, as in \citet{Nugroho20b}. There are a few possible explanations for this. Our shared models could have (1) overestimated the abundance of bound-free H-, (2) underestimated the temperature of the atmosphere, or (3) underestimated the strength of the thermal inversion. In future works, it would be worthwhile to further investigate the temperature structure and continuum opacity by varying these parameters in our models, but for now we opt to vary only the VMR, since varying additional parameters is computationally expensive. Though the effect is limited, a variable VMR still allows us to probe different temperatures as the atmosphere becomes optically thick at different pressures.


\subsection{Impact of T-P Profile} \label{sec:alt_analysis}
As a first step towards determining the impact of the temperature structure on our parameter constraints, we performed another analysis of our data using models with an increased stratospheric temperature. We did this by adjusting the ratio of the mean optical to infrared opacities substantially from from 2 to 5.5. We present this second analysis to investigate the impact of the T-P profile on high-resolution spectroscopic observations.

By cross-correlating our data with models using this alternative T-P profile, we found that the Fe {\sc i} signal was in fact detectable at blue wavelengths in the ESPaDOnS data. The increased stratospheric temperature produces a stronger temperature inversion, and as a result the line contrast of our model below $6000~\AA$ increases, impacting the detected signal. This is in line with our explanation in Section \ref{sec:bluewaves}. Therefore, in the following alternative analysis of our ESPaDOnS data, we include all wavelengths from $4000 - 9225 ~\AA$, only excluding the first and last four spectral orders due to noisy data at these outer edges \citep[see][for details]{Herman20}.

The likelihood results of this second analysis are presented in the last column of Table \ref{tab:results_data}, labeled `T-P Profile 2'. We found that, for all of our data sets, this stronger inversion had no serious impact on the phase offset, day-night contrast, or velocities. Only the $\alpha$ and $\log_{10}$VMR parameters were affected significantly, as expected. In our combined data set, for instance, the Fe {\sc i} VMR decreases by a factor of 20, while the $\alpha$ value suggests that these new models better estimate the Fe {\sc i} line contrast. However, the significance of our detection, based on the marginalized likelihood of $\alpha$, has decreased slightly. The day-night contrast and peak phase offset, on the other hand, are quite consistent between the two T-P profiles. Our ability to recover these parameters is further explored in our simulations presented in Section \ref{sec:sims}.

This presents a cautionary tale for high-resolution emission spectroscopy, as the abundances derived are affected by the T-P profile used, as illustrated by the differences in $\alpha$ and VMR between the two models. A full atmospheric retrieval, which is beyond the scope of this paper, would be required to address this in more detail. On a similar note, in some cases it may be valuable to instead fix the VMR and explore a variable lapse rate or T-P profile instead. However (with the exception of the VMR and $\alpha$), our fitted parameters are not greatly influenced by a change to the T-P profile, and therefore fitting for the VMR does not seem to affect our results in a significant way. But it would certainly be beneficial for future works to examine the T-P profile in greater detail.


\begin{deluxetable*}{lccccc}
\tabletypesize{\footnotesize}
\tablecaption{Summary of Simulated Spectra
\label{tab:sims}}
\tablehead{%
    \colhead{Sim. Name} & \colhead{Instrument} & \colhead{Orbital Phase} & \colhead{Cadence} & \colhead{No. Frames} & \colhead{SNR} \\
    &  & & \colhead{(min)} &  &  
    }
\startdata
ESP$_{\rm obs}$ & ESPaDOnS & $0.3-0.69$ & 2 & 440 & 245 \\
HDS$_{\rm obs}$ & HDS & $0.21-0.54$ & 10 & 52 & 330 \\
Comb. & ESPaDOnS \& HDS & $0.21 - 0.69$ & 2 \& 10 & 492 & 410 \\
ESP$_{250}$ & ESPaDOnS & $0.2-0.8$ & 2 & 526 & 250 \\
ESP$_{500}$ & ESPaDOnS & $0.2-0.8$ & 2 & 526 & 500 \\
ESP$_{750}$ & ESPaDOnS & $0.2-0.8$ & 2 & 526 & 750 \\
HDS$_{500}$ & HDS & $0.2-0.8$ & 2 & 526 & 500 \\
HDS$_{750}$ & HDS & $0.2-0.8$ & 2 & 526 & 750
\enddata
\end{deluxetable*}

\section{Investigating the Recovery of Phase Variations through Simulations}\label{sec:sims}
In this section, we produce simulated spectra with higher SNRs and greater phase coverage to determine how well the day-night contrast and peak phase offset can be constrained with high-resolution spectroscopy. We perform the same cross-correlation and likelihood analysis on eight sets of simulated data, the details of which are listed in Table \ref{tab:sims}. Since we are most interested in the effects on the day-night contrast and peak phase offset, and we have shown that the T-P profile has little impact on these brightness variation parameters, in these simulations we employ the atmospheric models used in our main analysis, rather than the models with an alternative T-P profile from Section \ref{sec:alt_analysis}.

\begin{figure*}
	\centering
    \includegraphics[width=0.66\textwidth]{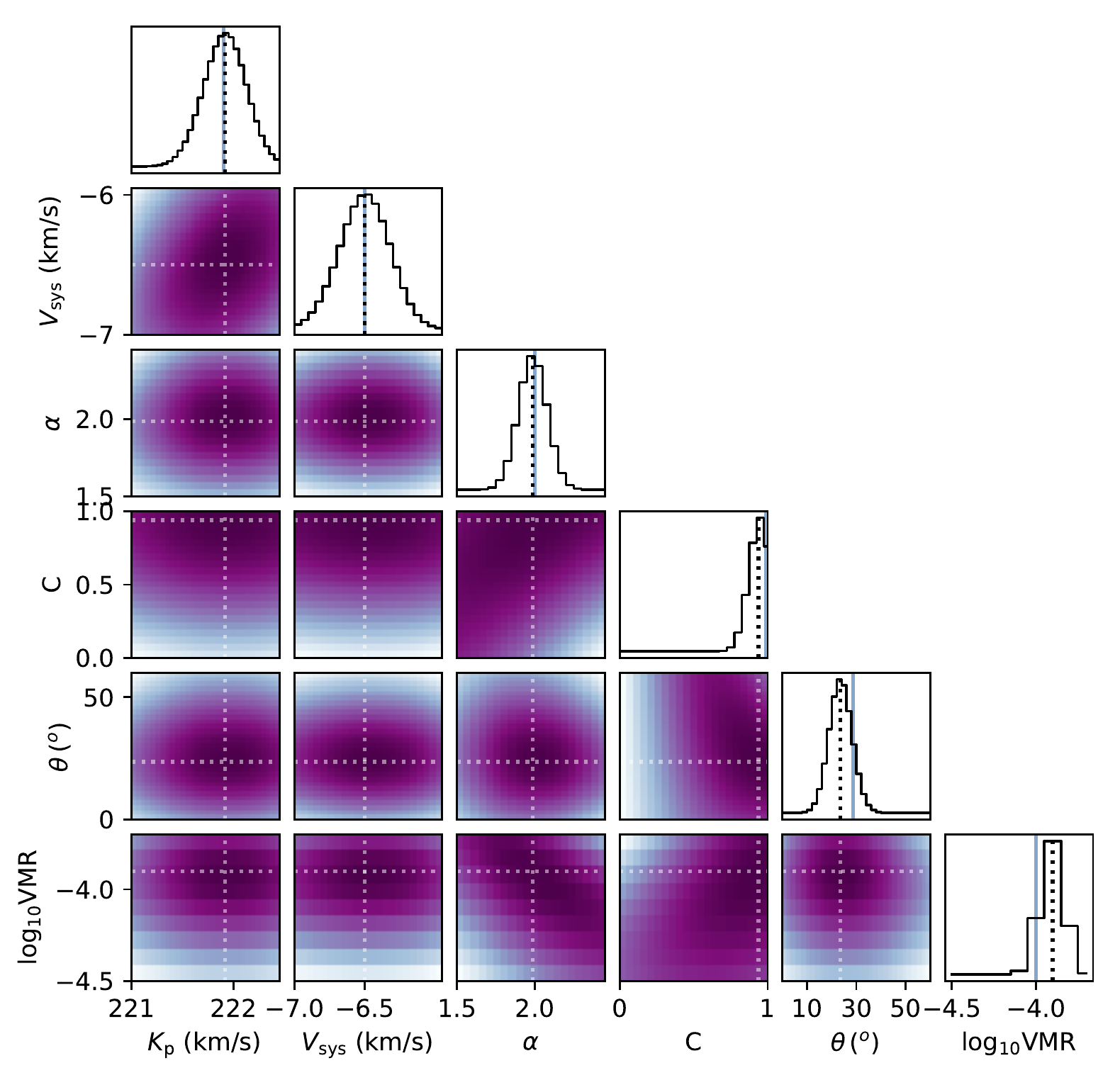}
    \caption{ The conditional (2D) and marginalized (1D) likelihood distributions for our simulation ESP$_{750}$, with an injected phase offset of $28.7^o$. The dotted lines shown the median values for each distribution, with the exception of $C$, which shows the lower limit estimated from the 84th percentile, and $\log_{10}$VMR, which shows the peak, since these distributions are non-Gaussian. The solid lines indicate the injected value for each parameter.
    }
    \label{fig:cornerplot_sim}
\end{figure*}

\subsection{Creating Simulated Spectra}
For the first and second simulations listed in the table, we first construct white noise (with a mean of 1) with the same average standard deviation along the time axis at each wavelength in our reduced data, and the same wavelength range as each spectral order in our reduced data. We do this for each night of ESPaDonS and HDS observations individually, using the same phase coverage and cadence for each night as listed in Table \ref{tab:obs}. These simulated data imitate the average noise properties of our observations (since the standard deviation is calculated from our detrended data), but exclude individual contaminating features from stellar and telluric lines, which are not perfectly removed by SYSREM. In general, our use of white noise is largely meant to remove any dependence on the effects of a single night (i.e., poor seeing, weather, etc.), to provide an example of what constraints may be possible for a typical set of observations with a given SNR and phase range.

The third simulation simply combines the data sets from the first two. For the remaining simulations, we create artificial spectra that cover a larger phase range of $0.2 - 0.8$ with a cadence of 2 minutes, to show the constraints we may place with observations with SNRs of 250, 500, and 750. The SNR of our original observations is estimated via $1/\langle\sigma\rangle$, where $\langle\sigma\rangle$ is the average standard deviation along the time axis in a $50~\AA$ window around $6825~\AA$. We chose this region because it is virtually devoid of stellar and telluric features. From here, we determine the scale factor by which the standard deviation can be multiplied to produce the desired SNR in our white noise simulations, taking into account the increased number of exposures. For the fourth through sixth simulations, we use as our baseline the standard deviation and estimated SNR of night 1 of our ESPaDOnS observations. For the last two simulations, we use the standard deviation and estimated SNR of our HDS observations. We then produce white noise for each case, with the appropriately scaled standard deviation along the time axis and wavelength range of each spectral order. 

Next, for all simulations we prepare a model for injection in the same manner as Section \ref{sec:modelprep}. We convolve the $\log_{10}{\rm VMR}=-4.0$ model to the resolution of the instrument, and apply a high-pass Butterworth filter. We then Doppler shift this model to the planetary velocity ($K_{\rm p} = 222$ km/s, $v_{\rm sys}=-6.5$ km/s) for each phase, and interpolate to the wavelength grid of each data set before multiplying the model into the white noise. The multiplied term is $1 + A_{\rm p} \times F_p/F_s$, where $A_{\rm p}$ is the brightness variation from equation \ref{eq:Ap}. We use the parameters $\alpha=2$, $C=0.99$, and consider three different phase offsets: $\theta = 0^o$, $-12.8^o$ \citep[from][]{Zhang18}, and $+28.7^o$ \citep[from][]{vonEssen20}. Thus we have a total of 24 simulated sets of spectra.

Finally, after the injection we subtract 1 from the simulated data and perform the same cross-correlation and likelihood mapping as before. Though our simulated data do not retain the stellar pulsations that contaminate our original cross-correlations, we still exclude orbital phases $0.41 - 0.59$ to replicate our original analysis as closely as possible. We also limit our simulated spectra to wavelengths $> 6000~\AA$, based on the findings presented in Section \ref{sec:bluewaves}.

\subsection{Results from Simulations} \label{sec:sims_results}
In Figure \ref{fig:cornerplot_sim} we show the marginalized and conditional distributions resulting from our analysis of simulation ESP$_{750}$ with an injected phase offset of $+28.7^o$, as an example. These spectra have the resolution and wavelength range of ESPaDOnS, with a 2-minute cadence over phases $0.2 - 0.8$, and a SNR of 750. Based on these results and those of the remaining simulations (which are too numerous to plot in the same manner), we find that the $K_{\rm p}$, $v_{\rm sys}$, $\alpha$, and $\log_{10}$VMR parameters are well constrained relative to their injected values. We further compared the best-fit parameters from our ESP$_{\rm obs}$, HDS$_{\rm obs}$, and Combined simulations to those we report for our real observations, respectively. We found that in all cases the constraints from our simulations, and particularly the uncertainties on each parameter, are generally consistent with the constraints from our real spectra. This suggests that our white noise is a sufficient approximation of the noise in our observations. 

\begin{figure}
	\centering
    \includegraphics[width=0.47\textwidth]{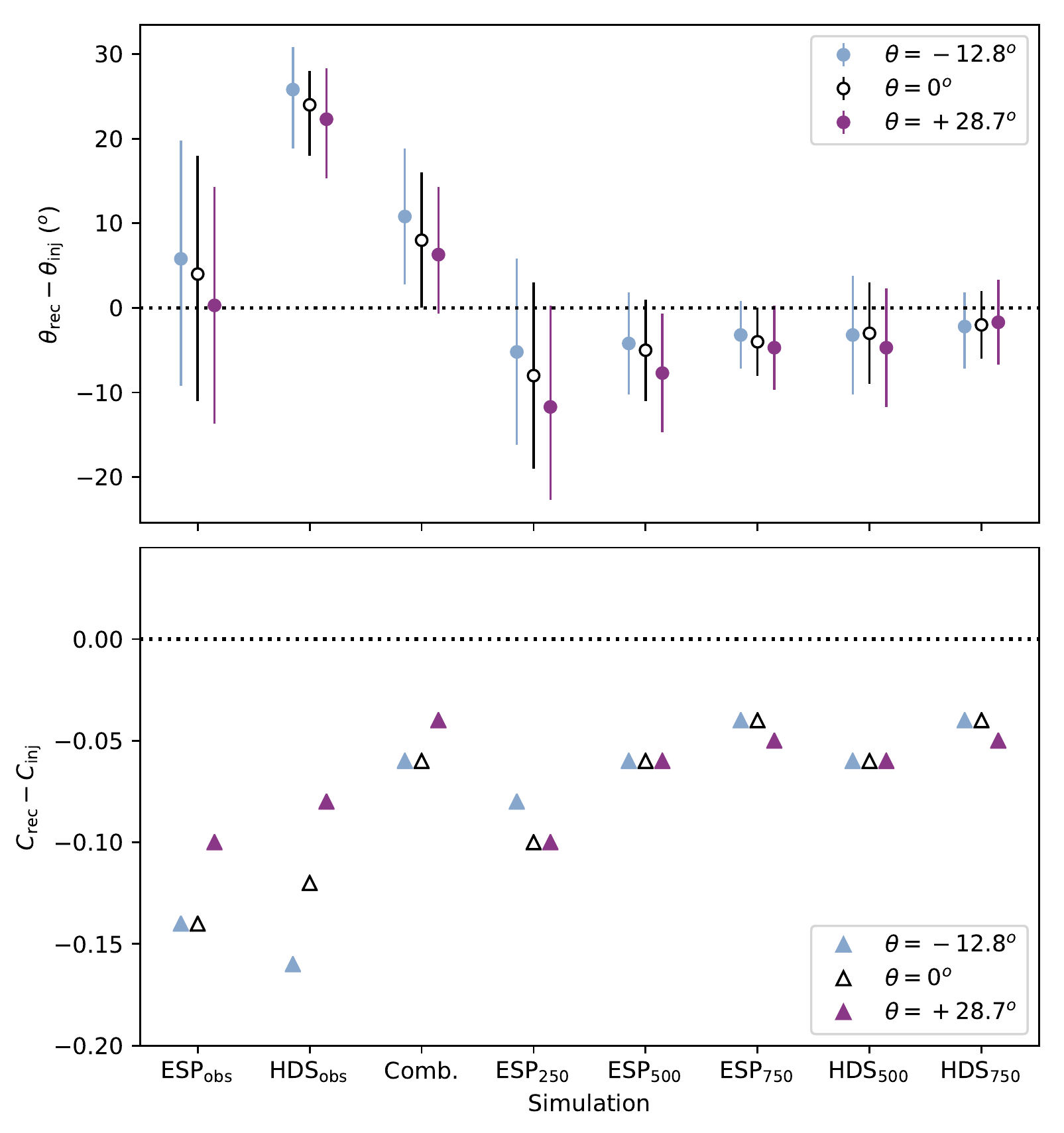}
    \caption{Comparison of our recovered phase offset (top) and day-night contrast (bottom) to their injected values for each simulation, plotted as the difference between the recovered and injected values. For the phase offset we use the median of the marginalized distribution $\pm1\sigma$, while for the contrast we use only the lower limit, taken as the 84th percentile of the bounded distribution.
    On the x-axis we label the simulations as they appear in Table \ref{tab:sims}. The light blue, white, and dark purple points show the results for injected models with phase offsets of $\theta= -12.8^o$, $0^o$, and $+28.7^o$, respectively. The injected contrast is kept at $C=0.99$ in all cases.
    }
    \label{fig:sim_comparison}
\end{figure}

Since we are most interested in the feasibility of accurately constraining the brightness variation terms with HRS, we therefore limit our remaining discussion to the recovery of the day-night contrast and phase offset parameters. Figure \ref{fig:sim_comparison} shows a comparison of our recovered phase offset and day-night contrast with their respective injected values for each simulation. As in Section \ref{sec:results_data}, we report the recovered phase offset as the median of the marginalized distribution $\pm1\sigma$, while for the contrast we report the lower limit, taken as the 84th percentile of the distribution. Our results from the first set of simulations, ESP$_{\rm obs}$, indicate that the recovered phase offsets for all three injected values are quite accurate, albeit with large uncertainties. This lends significant credit to the phase offset we report using our actual ESPaDOnS observations. The lower limits on the day-night contrast from these ESP$_{\rm obs}$ simulations are also similar to the lower limit we report from our real spectra. We again stress that $C$ is hard-bounded at 1, which will bias values away from the injected value of $C=0.99$.

As expected from our analysis of the actual HDS observations, the results from our simulated HDS spectra (labeled HDS$_{\rm obs}$) overestimate the phase offset by more than $20^o$. We reiterate that this is because (1) these data only probe the day-side atmosphere on one side of secondary eclipse, and (2) we exclude the phases where the peak brightness would occur, in order to avoid stellar pulsations. These limitations make it difficult to constrain the peak phase offset, and exemplify the need for observations on both sides of secondary eclipse. The confidence interval of the phase offset derived from the HDS data (both in our observations and simulations) may also appear relatively high, but it is possible to achieve a relatively tight constraint while having a large systematic offset from the actual value. This is especially true if the peak phase offset is not contained within our phase range, as is the case here. 
The effect this overestimated phase offset has on the combined spectra (labeled Comb.) is also apparent in our simulation results, reflecting our results from the real observations: The more accurate constraint from simulation ESP$_{\rm obs}$ is slightly outweighed by that of simulation HDS$_{\rm obs}$, due to the higher SNR of the latter. Interestingly, the lower limit on the contrast from simulation HDS$_{\rm obs}$ appears to be dependent on the injected phase offset. This also makes sense for observations with limited phase coverage: An eastward offset is somewhat degenerate with a lower contrast if only phases $<0.5$ are observed, and vice versa.

To test the bias toward positive phase offsets in greater detail, we ran three additional sets of simulations based on the HDS data with injected offsets of $-35^o$, $0^o$, and $+28^o$, each with 25 different realizations of the noise. We kept the other parameters equal to the best fit model. In general, we found that the RMS between the different simulations was larger than the uncertainty we found for our data, but that the bias was always present. However, the magnitude of the bias (both in size and number of simulations) increased with an offset towards the planet's morning side, which would imply that the bias is due to the limited phase-coverage. From this we conclude that the bias we see in our results is likely real, but that we may be underestimating the uncertainties on the recovered offset.

In more detail, the average uncertainties on the phase offset obtained from our additional simulations were larger than those retrieved from our HDS observations (e.g., $\pm20^o - 30^o$ compared to $\pm6^o$). Given that an increase in uncertainty could change the nature of any detected phase offset from significant to non-significant, this warrants a closer look. The discrepancy between the phase offset uncertainty from our additional simulations and that of our observations is, at least in part, driven by the aforementioned points that bias the results. Furthermore, the increased scatter in our additional simulations appeared to be driven by a few simulations with different noise realizations where a higher value of $\theta$ was preferred. We also note that in several cases, the fits hit the upper/lower boundary of the parameter grid, which can skew the results slightly, and due to the extensive run time/memory requirements of these simulations, it was not feasible to extend the parameters to a larger range. All of these effects could influence the uncertainty on the phase offset in our additional HDS simulations.

With that said the vast majority of the weight in our combined results, in both our original simulations and observations, originates from the HDS data. Our ESPaDOnS results should have a smaller bias due to coverage on both sides of the day-side phase curve, and therefore the uncertainties on this retrieved phase offset should be less affected.

Continuing with our results in Figure \ref{fig:sim_comparison}, the remaining points show the results for simulations with higher SNRs and more complete day-side phase coverage. These show that extended observations on both sides of secondary eclipse can greatly improve the accuracy of the recovered contrast and phase offset, particularly compared to the HDS$_{\rm obs}$ set of simulations. Observations with higher SNRs also provide tighter constraints on both the phase offset and day-night contrast. The HDS simulations provide slightly better constraints on the phase offset than the ESPaDOnS simulations at the same SNRs, which is due to the higher resolution of HDS. 

Overall, our HRS observations and simulations both show that, given sufficient phase coverage with high SNR spectra, it is entirely feasible to constrain the day-night contrast and phase offset of a hot Jupiter atmosphere. Such constraints can inform aspects of energy re-circulation in a planet's atmosphere, though the avoidance of stellar pulsations in the CCF is a necessity. If such pulsations can be removed from the CCF around secondary eclipse, we may be able to even better constrain the peak phase offset, because the phase at which it occurs will no longer need to be excluded. This is a complex issue that is outside the scope of this work. Though our simulated spectra do not feature stellar pulsations, we still choose to exclude the affected phases from our analysis, since the feasibility of removing pulsation signals from emission spectra remains an open question. Pulsation signals have already been successfully removed from transmission spectra \citep[e.g.,][though the planetary signal in the latter was artificially injected]{Johnson15,Temple17,vanSluijs19}, but adapting these methods to day-side emission spectra will not be trivial. However, efforts along these lines will be invaluable for future atmospheric investigations of planets orbiting variable stars, such as WASP-33b, $\beta$ Pic b, and WASP-176b/KELT-13b.

\section{Conclusions}\label{sec:conclusion}

We have detected Fe {\sc i} in the atmosphere of the ultra-hot Jupiter WASP-33b, indicative of a thermal inversion layer produced by an atomic species. Combining optical high-resolution data sets from five nights of observations with ESPaDOnS on CFHT and one night with HDS on Subaru,  we make a detection of $>10.4\sigma$. By employing a likelihood mapping technique, we are able to constrain the RV semi-amplitude to $K_{\rm p} = 222.1_{-0.5}^{+0.6}$ km/s and the systemic velocity to $v_{\rm sys} = -6.5\pm0.3$ km/s, matching previously reported values. We find that the Fe {\sc i} signal is only detected above $6000~\AA$ in our ESPaDOnS spectra, and use simulated data to explain that this phenomenon is due to decreased line contrast at bluer wavelengths. By cross-correlating with models with a higher stratospheric temperature, and therefore greater line contrast, we further show that a signal at these wavelengths can actually be detected.

We also introduce a brightness variation term to our likelihood calculations, to explore the day-night contrast and phase offset of the peak brightness from our individual and combined data sets, assuming the planetary emission spectrum is dominated by Fe {\sc i} lines. By performing the same analysis on simulated spectra, we then explore how well these parameters can be constrained from observations with higher SNRs and greater phase coverage. We find that our constraint on the phase offset from HDS is likely overestimated due to the limited phase coverage of those observations, and we show that this further affects the constraint from our combined data set, since the SNR of our HDS observations is much higher than that of our ESPaDOnS observations. We therefore suggest that our reported constraint of $\theta=+22\pm12$ degrees from ESPaDOnS alone is likely a better estimate of the true phase offset of WASP-33b, since these observations cover phases before and after secondary eclipse relatively evenly. We report a day-night contrast of $C > 0.90$ using the same spectra, indicating that the night-side flux is $<10\%$ of the day-side flux in the wavelength range we consider. Using atmospheric models with a stronger temperature inversion, we also show that these brightness variation parameters are not significantly impacted by the T-P profile selected. Given the dependence of our models on the assumed lapse rate, however, our results cannot be directly used to infer the absolute abundance and metallicity of the planet's atmosphere. A detailed interpretation of the day-night contrast would also need to take into account a variable lapse rate as a function of position on the planet.

Based on both our observations and simulations, high-resolution spectra that cover a large phase range on both sides of secondary eclipse can be used to constrain a planet's day-night contrast and peak phase offset, even in the presence of stellar variability. This approach provides a new means to confirm or challenge existing measurements from photometric light curves. We encourage further  investigation into the removal of stellar pulsation signals from spectroscopic data; this has been done in a few cases \citep[e.g.,][]{Johnson15, Temple17, vanSluijs19}, but adapting the procedure to day-side emission observations will present unique challenges. We also encourage future studies to examine more complete models of a planet's brightness variation throughout its orbit. They would facilitate a deeper exploration of energy circulation within a planetary atmosphere, and expand the utility of high-resolution Doppler spectroscopy in characterizing exoplanets.


\section*{Acknowledgements}
Some of the data presented herein were obtained at the Canada-France-Hawaii Telescope, which is operated by the National Research Council of Canada, the Institut National des Sciences de l'Univers of the Centre National de la Recherche Scientique of France, and the University of Hawaii. Our second data set was collected at the Subaru Telescope, which is operated by the National Astronomical Observatory of Japan. The authors also wish to recognize and acknowledge the very significant cultural role and reverence that the summit of Maunakea has had within the indigenous Hawaiian community.  We are most fortunate to have the opportunity to conduct observations from this mountain. M.K.H. was supported by funding from the Natural Sciences and Engineering Research Council (NSERC) of Canada.


\appendix

\section{Model Filtering}\label{sec:model_filt}

As mentioned in Section \ref{sec:modelprep}, we apply a Butterworth filter to each model prior to cross-correlating, to mimic the effects of SYSREM on the planetary signal. Each night of observations has unique noise properties, and requires a different number of SYSREM iterations to account for the stellar and telluric features present. The filter parameters that best represent the effects of SYSREM will then be different for each night, and to determine those parameters, we perform a series of injection and recovery tests.

For each injection, we use the same models described in Section \ref{sec:models} but do not apply the Butterworth filter. The model is injected into the extracted spectra (i.e., before blaze correction, SYSREM, etc.) at the same orbital speed as the real Fe signal, but with the opposite sign, so that the real and injected planetary signals do not overlap, but are treated similarly in our data reduction procedure. The model is Doppler shifted to the planet's velocity for each frame, interpolated to the wavelength grid of each spectral order, and multiplied into the data in the form $1 + F_p/F_s$. We then apply the same blaze correction and number of SYSREM iterations as are used for each night (Table \ref{tab:obs}).

\begin{figure}
	\centering
    \includegraphics[width=0.47\textwidth]{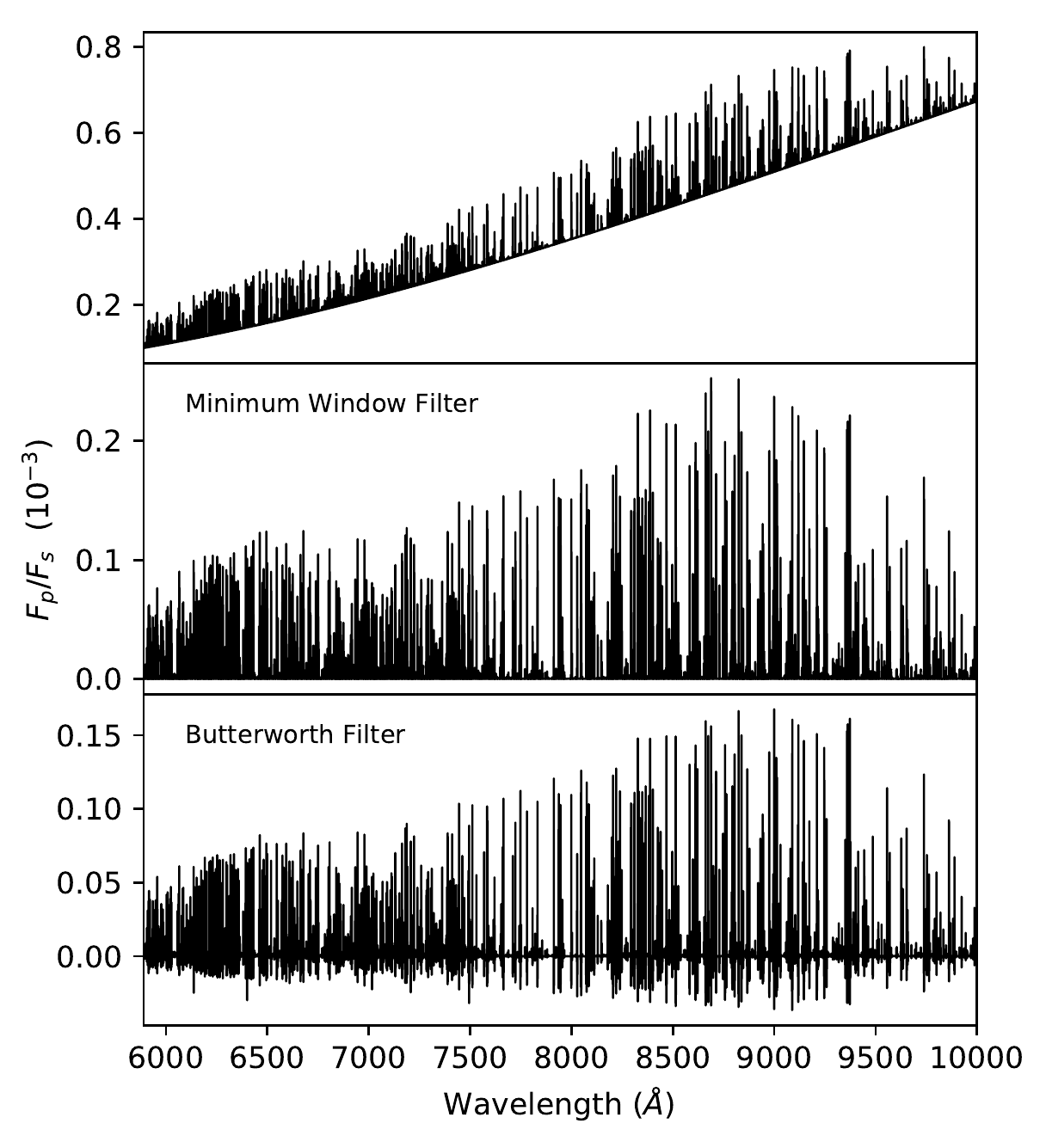}
    \caption{An example of the planetary emission model with $\log_{10}{\rm VMR} = -4$, to which different filtering methods have been applied. In the top panel we have only convolved the model to the resolution of the ESPaDOnS data and calculated the planet-star flux ratio. In the center panel we have additionally applied a minimum window filter with a window of 4 $\AA$ to remove the planetary continuum. In the bottom panel we have instead applied a high-pass Butterworth filter to the model from the top panel, which better accounts for the effect of SYSREM on the planetary signal.
    }
    \label{fig:model_filter}
\end{figure}

Next, we perform the same cross-correlation and likelihood computation as before, this time searching for the injected signal, and fixing the day-night contrast and phase offset to zero. The model with which we cross-correlate (after convolving to the instrument resolution and calculating the planet-star flux ratio) is initially filtered using a minimum window filter with a window of $\sim4~\AA$, to subtract only the planetary continuum from the model. This filtering method is used often with Doppler cross-correlation \citep[e.g.,][]{Herman20, Nugroho20b, Jindal20}. However, it does not account for the effects that the SYSREM algorithm has on the injected planetary signal, and as a result, the best-fitting $\alpha$ parameter underestimates the true scale of the injected model -- generally by $\sim30$\% or more. This has clear consequences: if we do not address the side effects of SYSREM when dealing with the real planetary signal, the true nature of the underlying emission spectrum cannot be properly understood.

We therefore follow the example of \citet{Gibson20}, and introduce a high-pass Butterworth filter in place of the minimum window filter. We then adjust the parameters of the filter for each night (i.e., the filter order and cut-off frequency). We find that it is possible to recover a value for $\alpha$ much closer to the true scale of the injection through more aggressive filtering of the model with which we cross-correlate. For our various data sets, this means increasing the cut-off frequency until the recovered $\alpha$ is within a few percent of the injected value. In Figure \ref{fig:model_filter}, we show an example model before any filtering, after applying a minimum window filter, and after applying a Butterworth filter.

Initially, we determined the optimal filter parameters for a given night using a single model for the injection and cross-correlation. However, we confirmed that the chosen parameters were also appropriate for models with different VMRs; in other words the filter order and cut-off frequency were not optimized to recover a specific model. Additionally, this filtering did not have any noticeable impact on the recovery of other injected parameters. 

Through these injection tests, we are able to determine the appropriate filter parameters for our models when cross-correlating with the actual planetary spectra for each night, so that the best-fitting $\alpha$ value is properly estimated. Of course, this singular filtering method cannot perfectly reproduce the complex effects of SYSREM on the planetary signal. We simply consider this to be a first step towards addressing the nuances commonly overlooked in Doppler cross-correlation analyses. As the analysis techniques developed for HRS become increasingly advanced, our data processing methods must evolve as well. Looking forward, an in-depth study of the implications of our treatment of high-resolution data would be well worth the time investment.


\section{Additional Figures}\label{sec:extra}

\begin{figure*}
	\centering
    \includegraphics[width=0.66\textwidth]{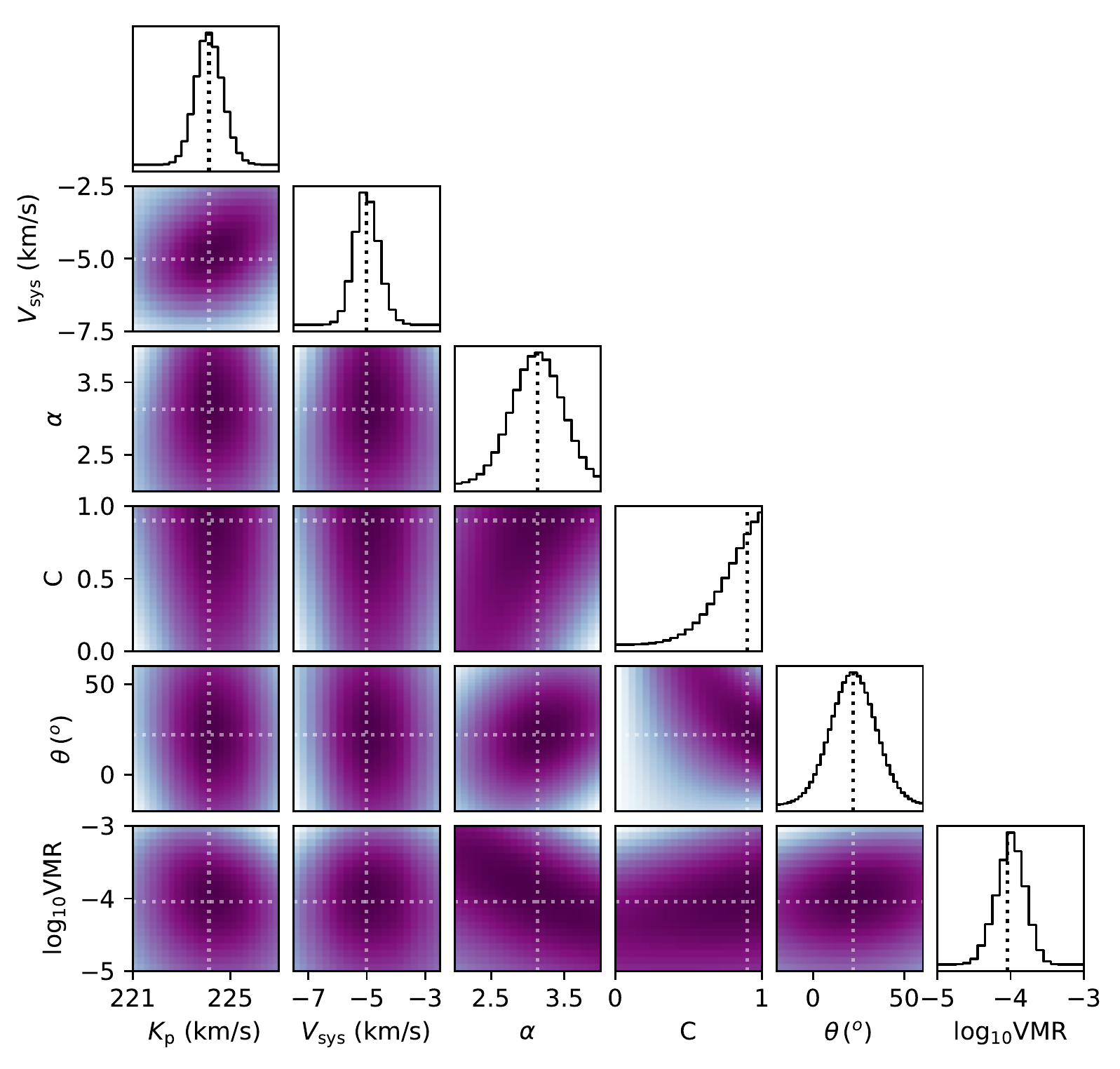}
    \caption{The same as Figure \ref{fig:cornerplot_data}, but for the likelihood distributions of our ESPaDOnS observations alone. Note that the axis ranges of some subplots differ between these figures.
    }
    \label{fig:cornerplot_cfht}
\end{figure*}

\begin{figure*}
	\centering
    \includegraphics[width=0.66\textwidth]{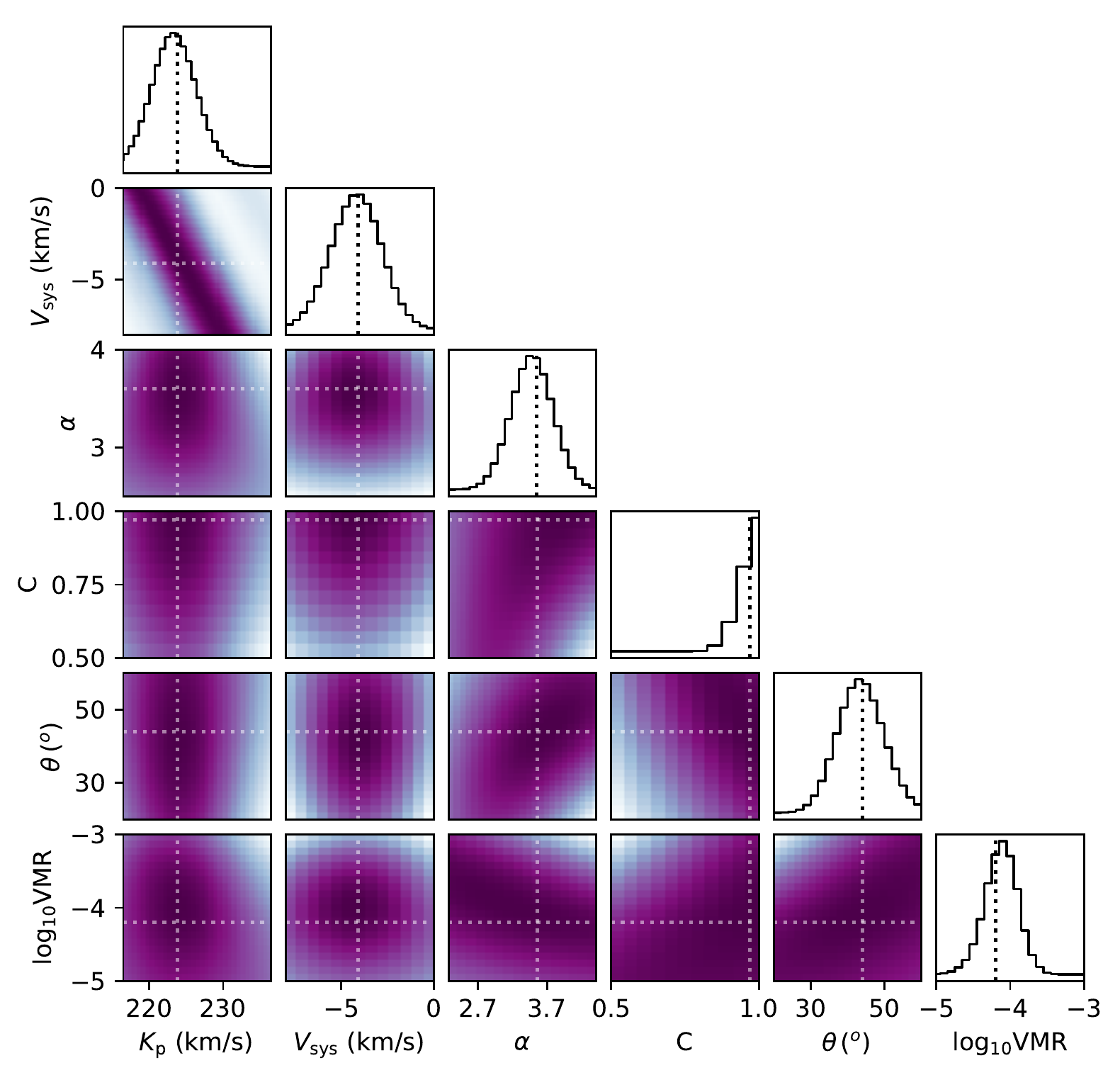}
    \caption{The same as Figures \ref{fig:cornerplot_data} and \ref{fig:cornerplot_cfht}, but for the likelihood distributions of our HDS observations alone.
    }
    \label{fig:cornerplot_hds}
\end{figure*}

In Figures \ref{fig:cornerplot_cfht} and \ref{fig:cornerplot_hds}, we show the same marginalized and conditional likelihood distributions as Figure \ref{fig:cornerplot_data}, but for our individual ESPaDOnS and HDS results, respectively.

\bibliography{bibliography}{}
\bibliographystyle{aasjournal}

\end{document}